\font\tenfrakturb=eufb10
\font\tenfraktur=eufm10
\font\tenmsbm=msbm10
\font\sevenfrakturb=eufb7
\font\sevenfraktur=eufm7
\font\sevenmsbm=msbm7
\font\fivefrakturb=eufb5
\font\fivefraktur=eufm5
\font\fivemsbm=msbm5
\newfam\bgothicfam
\newfam\gothicfam
\newfam\msbmfam
\textfont\bgothicfam = \tenfrakturb \scriptfont\bgothicfam=\sevenfrakturb
\scriptscriptfont\bgothicfam=\fivefrakturb
\textfont\gothicfam = \tenfraktur \scriptfont\gothicfam=\sevenfraktur
\scriptscriptfont\gothicfam=\fivefraktur
\textfont\msbmfam = \tenmsbm \scriptfont\msbmfam=\sevenmsbm
\scriptscriptfont\msbmfam=\fivemsbm

\def\Bbb{\tenmsbm\fam\msbmfam}

\catcode`@=11
\def\renewcounter#1{\@definecounter{#1}\@ifnextchar[{\@newctr{#1}}{}}
\documentstyle[twoside]{article}
\catcode`\@=11
\long\def\@makefntext#1{
\protect\noindent \hbox to 3.2pt {\hskip-.9pt  
$^{{\eightrm\@thefnmark}}$\hfil}#1\hfill}		

\def\@makefnmark{\hbox to 0pt{$^{\@thefnmark}$\hss}}	
	
\def\ps@myheadings{\let\@mkboth\@gobbletwo
\def\@oddhead{\hbox{}
\rightmark\hfil\eightrm\thepage}   
\def\@oddfoot{}\def\@evenhead{\eightrm\thepage\hfil
\leftmark\hbox{}}\def\@evenfoot{}
\def\sectionmark##1{}\def\subsectionmark##1{}}
\oddsidemargin=\evensidemargin
\newcounter{sectionc}\newcounter{subsectionc}\newcounter{subsubsectionc}
\renewcommand{\section}[1] {\vspace{12pt}\addtocounter{sectionc}{1} 
\setcounter{subsectionc}{0}\setcounter{subsubsectionc}{0}\noindent 
{\tenbf\thesectionc. #1}\par\vspace{5pt}}
\renewcommand{\subsection}[1] {\vspace{12pt}\addtocounter{subsectionc}{1} 
\setcounter{subsubsectionc}{0}\noindent 
{\bf\thesectionc.\thesubsectionc. {\kern1pt \bfit #1}}\par\vspace{5pt}}
\renewcommand{\subsubsection}[1]{\vspace{12pt}\addtocounter{subsubsectionc}{1}
\noindent{\tenrm\thesectionc.\thesubsectionc.\thesubsubsectionc.
{\kern1pt \tenit #1}}\par\vspace{5pt}}
\newcommand{\nonumsection}[1] {\vspace{12pt}\noindent{\tenbf #1}
\par\vspace{5pt}}
\newcounter{appendixc}
\newcounter{subappendixc}[appendixc]
\newcounter{subsubappendixc}[subappendixc]
\renewcommand{\thesubappendixc}{\Alph{appendixc}.\arabic{subappendixc}}
\renewcommand{\thesubsubappendixc}
{\Alph{appendixc}.\arabic{subappendixc}.\arabic{subsubappendixc}}
\renewcommand{\appendix}[1] {\vspace{12pt}
        \refstepcounter{appendixc}
        \setcounter{figure}{0}
        \setcounter{table}{0}
        \setcounter{lemma}{0}
        \setcounter{theorem}{0}
        \setcounter{corollary}{0}
        \setcounter{definition}{0}
        \setcounter{equation}{0}
        \renewcommand{\thefigure}{\Alph{appendixc}.\arabic{figure}}
        \renewcommand{\thetable}{\Alph{appendixc}.\arabic{table}}
        \renewcommand{\theappendixc}{\Alph{appendixc}}
        \renewcommand{\thelemma}{\Alph{appendixc}.\arabic{lemma}}
        \renewcommand{\thetheorem}{\Alph{appendixc}.\arabic{theorem}}
        \renewcommand{\thedefinition}{\Alph{appendixc}.\arabic{definition}}
        \renewcommand{\thecorollary}{\Alph{appendixc}.\arabic{corollary}}
        \renewcommand{\theequation}{\Alph{appendixc}.\arabic{equation}}
        \noindent{\tenbf Appendix \theappendixc #1}\par\vspace{5pt}}
\newcommand{\subappendix}[1] {\vspace{12pt}
        \refstepcounter{subappendixc}
        \noindent{\bf Appendix \thesubappendixc. {\kern1pt \bfit #1}}
\par\vspace{5pt}}
\newcommand{\subsubappendix}[1] {\vspace{12pt}
        \refstepcounter{subsubappendixc}
        \noindent{\rm Appendix \thesubsubappendixc. {\kern1pt \tenit #1}}
\par\vspace{5pt}}
\newcommand{\textlineskip}{\baselineskip=13pt}
\newcommand{\smalllineskip}{\baselineskip=10pt}
\def\eightcirc{
\begin{picture}(0,0)
\put(4.4,1.8){\circle{6.5}}
\end{picture}}
\def\eightcopyright{\eightcirc\kern2.7pt\hbox{\eightrm c}} 
\newcommand{\copyrightheading}[1]
{\vspace*{-2.5cm}\smalllineskip{\flushleft
{\footnotesize International Journal of Modern Physics A, #1}\\
{\footnotesize $\eightcopyright$\, World Scientific Publishing
 Company}\\
 }}

\newcommand{\pub}[1]{{\begin{center}\footnotesize\smalllineskip 
Received #1\\
\end{center}
}}

\def\abstracts#1#2#3{{
\centering{\begin{minipage}{4.5in}\baselineskip=10pt\footnotesize
\parindent=0pt #1\par 
\parindent=15pt #2\par
\parindent=15pt #3
\end{minipage}}\par}} 
\def\keywords#1{{
\centering{\begin{minipage}{4.5in}\baselineskip=10pt\footnotesize
{\footnotesize\it Keywords}\/: #1
 \end{minipage}}\par}}

\renewenvironment{thebibliography}[1]
{\frenchspacing
 \ninerm\baselineskip=11pt
 \begin{list}{\arabic{enumi}.}
{\usecounter{enumi}\setlength{\parsep}{0pt}
 \setlength{\leftmargin 12.7pt}{\rightmargin 0pt} 
 \setlength{\itemsep}{0pt} \settowidth
{\labelwidth}{#1.}\sloppy}}{\end{list}}
\newcounter{itemlistc}
\newcounter{romanlistc}
\newcounter{alphlistc}
\newcounter{arabiclistc}

\newcommand{\fcaption}[1]{
        \refstepcounter{figure}
        \setbox\@tempboxa = \hbox{\footnotesize Fig.~\thefigure. #1}
        \ifdim \wd\@tempboxa > 5in
           {\begin{center}
        \parbox{5in}{\footnotesize\smalllineskip Fig.~\thefigure. #1}
            \end{center}}
        \else
             {\begin{center}
             {\footnotesize Fig.~\thefigure. #1}
              \end{center}}
        \fi}
\newcommand{\tcaption}[1]{
        \refstepcounter{table}
        \setbox\@tempboxa = \hbox{\footnotesize Table~\thetable. #1}
        \ifdim \wd\@tempboxa > 5in
           {\begin{center}
        \parbox{5in}{\footnotesize\smalllineskip Table~\thetable. #1}
             \end{center}}
        \else
             {\begin{center}
             {\footnotesize Table~\thetable. #1}
              \end{center}}
        \fi}
\def\@citex[#1]#2{\if@filesw\immediate\write\@auxout
{\string\citation{#2}}\fi
\def\@citea{}\@cite{\@for\@citeb:=#2\do
{\@citea\def\@citea{,}\@ifundefined
{b@\@citeb}{{\bf ?}\@warning
{Citation `\@citeb' on page \thepage \space undefined}}
{\csname b@\@citeb\endcsname}}}{#1}}
\newif\if@cghi
\def\cite{\@cghitrue\@ifnextchar [{\@tempswatrue
\@citex}{\@tempswafalse\@citex[]}}
\def\citelow{\@cghifalse\@ifnextchar [{\@tempswatrue
\@citex}{\@tempswafalse\@citex[]}}
\def\@cite#1#2{{$\null^{#1}$\if@tempswa\typeout
{IJCGA warning: optional citation argument 
ignored: `#2'} \fi}}

\def\pmb#1{\setbox0=\hbox{#1}
\kern-.025em\copy0\kern-\wd0
\kern.05em\copy0\kern-\wd0
\kern-.025em\raise.0433em\box0}

\def\fnt#1#2{\footnotetext{\kern-.3em
{$^{\mbox{\scriptsize #1}}$}{#2}}}
\def\fpage#1{\begingroup
\voffset=.3in
\thispagestyle{empty}\begin{table}[b]\centerline{\footnotesize #1}
\end{table}\endgroup}
\def\runninghead#1#2{\pagestyle{myheadings}
\markboth{{\protect\footnotesize\it{\quad #1}}\hfill}
{\hfill{\protect\footnotesize\it{#2\quad}}}}
\font\tenrm=cmr10
\font\tenit=cmti10 
\font\tenbf=cmbx10
\font\bfit=cmbxti10 at 10pt
\font\ninerm=cmr9

\font\eightrm=cmr8

\def\bh{${\Bbb R}^2\times {\Bbb S}^2\>$}

\def\qed{\hbox{${\vcenter{\vbox{  
   \hrule height 0.4pt\hbox{\vrule width 0.4pt height 6pt
   \kern5pt\vrule width 0.4pt}\hrule height 0.4pt}}}$}}

\begin{document}
\runninghead{Yu. P. Goncharov \& N. E. Firsova}
{ Increase of the electron-positron
 Hawking radiation }
\normalsize\textlineskip
\thispagestyle{empty}
\setcounter{page}{761}
\copyrightheading{Vol. 19, No. 5 (2004) 761-774}
\vspace*{0.88truein}
\fpage{761}
\centerline{\bf INCREASE OF THE}
\vspace*{0.035truein}
\centerline{\bf ELECTRON-POSITRON HAWKING RADIATION}
\centerline{\bf FROM SCHWARZSCHILD BLACK HOLES}
\vspace*{0.035truein}
\centerline{\bf BY DIRAC MONOPOLES}
\vspace*{0.37truein}
\centerline{\footnotesize YU. P. GONCHAROV}
\vspace*{0.015truein}
\centerline{\footnotesize\it Theoretical Group,
Experimental Physics Department, State Polytechnical University}
\baselineskip=10pt
\centerline{\footnotesize\it Sankt-Petersburg 195251, Russia}
\vspace*{10pt} 
\centerline{\footnotesize N. E. FIRSOVA}
\vspace*{0.015truein}
\centerline{\footnotesize\it Institute for Problems of Mechanical 
Engineering}
\baselineskip=10pt
\centerline{\footnotesize\it Russian Academy of Sciences, Sankt-Petersburg 
199178, Russia}
\vspace*{10pt}
\vspace*{0.225truein}
\pub{18 November 2003}
\vspace*{0.21truein}
\abstracts{
 An algorithm for numerical computation of the barrier transparency for
the potentials surrounding Schwarzschild black holes is described for
massive spinor particles. It is then applied to calculate the total
(including electronic neutrino and the contributions of twisted field
configurations connected with Dirac monopoles)
luminosity for the electron-positron Hawking radiation from a Schwarzschild black hole
with mass $M=10^{15}$ g. It is found that the contribution due to monopoles
can be of order 12~\%  of the total electron-positron luminosity.
}{}{}
\vspace*{10pt}
\vspace*{1pt}\textlineskip 
\section{Introduction} 
\vspace*{-0.5pt}
\noindent

The present paper is a natural continuation of our works
\cite{{GF},{GF01}} so, referring for more details on motivation of the given 
trend to the mentioned references, we shall here restrict ourselves to a short 
discussion.

  Although it has elapsed over 25 years since Hawking discovered the 
possibility for black holes to radiate quantum particles, the situation with
quantitative study of the Hawking radiation up to recently should be considered
as extremely unsatisfactory. If qualitative understanding of the process
developed more or less successfully then attempts of quantitative calculations
might be enumerated by fingers. Really one may speak only about works of
Refs.\cite{Page} and those of Refs.\cite{Mac90} based on Page's ones. 
However, when analysing the mentioned papers one can note that central problem
of numerical calculations for Hawking radiation --  calculating the barrier 
transparency for the potentials surrounding black holes -- was actually not
resolved so it is impossible to extract some exact algorithm for numerical 
computation of the necessary quantities from the mentioned references.

  Another aspect of the problem in question is connected with that
any isolated black hole might possess the internal magnetic fields of the Dirac
monopole types. The latter configurations should be connected with nontrivial
topological properties of black holes and could have an essential influence on
quantum processes near black holes, for instance, on Hawking radiation. 
A number of examples of such configurations may be found in Refs.\cite{GF} and
references therein. Physically, the existence of those configurations should
be obliged to the natural presence of magnetic U(N)-monopoles (with $N\ge1$)
on black holes though the total (internal) magnetic charge (abelian or
nonabelian) of black hole remains equal to zero. One can consider that
monopoles reside in black holes as quantum objects without having influence on
the black hole metrics. They could reside in the form of monopole gas in which
the process of permanent creation and annihilation of the virtual
monopole-antimonopole pairs occurs so that the summed internal magnetic charge
(i. e., related with topological properties) is equal to zero while the external
one (not connected with topological properties) may differ from zero. While
existing the virtual monopole-antimonopole pair can interact with a particle
and, by this, increasing the Hawking radiation (see Refs.\cite{GF} and
references therein). 

  In other words one may say that the black holes due to
their nontrivial topological properties can actually carry the whole spectrum 
of topologically inequivalent configurations (TICs) for miscellaneous fields, 
in the first turn, complex scalar and spinor ones.
The mentioned TICs can markedly modify the Hawking radiation from black
holes. Physically, the existence of TICs should
be obliged to the natural presence of magnetic U(N)-monopoles (with $N\ge1$)
on black holes and additional contributions to the Hawking radiation exist due 
to the additional scalar or spinor particles leaving the black hole because 
of the interaction with monopoles so the conforming radiation can be called 
{\it the monopole Hawking radiation}.\cite{Gon99}

 Up to now, however, only influence of the TICs of complex scalar field 
on Hawking radiation has been studied more or less (see Refs.\cite{GF} and
references therein for more details). The description of TICs for
spinors was obtained in Refs.\cite{Gon99} but the detailed analysis of the
TICs contribution to Hawking radiation requires knowledge of the conforming
$S$-matrices which regulate the spinor particle passing through the potential
barrier surrounding black hole. Those $S$-matrices for the Schwarzschild (SW)
black holes have only recently been explored in Ref.\cite{Fir01} which allows 
us to obtain an algorithm to
calculate the $S$-matrix elements numerically since for physical results to
be obtained one needs to apply the numerical methods. In the massless case the
$S$-matrices discussed are simpler to treat and the paper of Ref.\cite{GF01}
contained a description of the algorithm needed and applied it to calculate 
the all configurations luminosity for massless spinor particles for a SW black 
hole.
The results obtained can serve as an estimate, in the first turn, of the
electron-positron Hawking radiation and also for the neutrino one from the
SW black holes. The more exact computation should take into account the
particle masses and require more complicated algorithms for calculating
the corresponding $S$-matrices.

The present paper contains a description of one such an algorithm and applies 
the latter to calculate the total luminosity for electrons, positrons, 
electronic neutrino and antineutrino for a SW black hole with mass 
$M=10^{15}$ g. Section 2 is devoted to the common statement of the problem
while Section 3 gives more detailed information about the relations
needed for further calculations. Section 4 estimates the correctness of 
computations and Section 5 presents the numerical results obtained. Finally,
Section 6 is devoted to concluding remarks.

We write down the black hole
metric under discussion (using the ordinary set of local coordinates
$t,r,\vartheta,\varphi$) in the form
$$ds^2=adt^2-a^{-1}dr^2-r^2(d\vartheta^2+\sin^2\vartheta d\varphi^2) 
\eqno(1)$$
with $a=1-2M/r$ and $M$ is the black hole mass.

  Throughout the paper we employ the system of units with $\hbar=c=G=1$,
unless explicitly stated otherwise. We shall denote $L_2(F)$ the set of the 
modulo square integrable
complex functions on any manifold $F$ furnished with an integration measure
while $L^n_2(F)$ will be the $n$-fold direct product of $L_2(F)$
endowed with the obvious scalar product.

Finally, when computing we use the following
value of the electron-positron mass (see Ref.\cite{pdg})
$$\mu_0(e^\pm)= 0.51099906\ {\rm Mev}\>.$$

\section{Preliminaries}
As was disscussed in Refs.\cite{Gon99},
TICs of a spinor field on black holes are conditioned by the
availability of a countable
number of the twisted spinor bundles over the \bh-topology underlying
the 4D black hole physics. From a physical point of view
the appearance of spinor twisted configurations is linked with the natural
presence of Dirac monopoles that play the role of connections in the
complex line bundles corresponding to the twisted spinor bundles.
Under the circumstances each TIC corresponds to sections of the corresponding
spinor bundle $E$, which can be characterized by its Chern number
$n\in \Bbb{Z}$
(the set of integers).
Using the fact that all the mentioned bundles can be trivilized over
the chart of local coordinates
$(t,r,\vartheta,\varphi) $
covering almost the whole manifold \bh
one can obtain a suitable Dirac equation on the given chart for TIC
$\Psi$
with mass $\mu_0$ and Chern number $n\in\Bbb{Z}$ that looks
as follows
$${\cal D}_n\Psi=\mu_0\Psi,\>\eqno(2)$$
with the twisted Dirac operator ${\cal D}_n=i\gamma^\mu\nabla_\mu^n$
and we can call (standard) spinors corresponding to $n=0$
{\it untwisted} while the rest of the spinors with $n\ne0$
should be referred to as {\it twisted}. Referring for details and for
explicit form of ${\cal D}_n$ to Refs.\cite{Gon99}, it should be noted
here that in $L_2^4$(\bh) there is a basis from the solutions of (2)
in the form
$$\Psi_{\lambda m}=\frac{1}{\sqrt{2\pi\omega}}
e^{i\omega t}r^{-1}\pmatrix{F_1(r,\omega,\lambda)
\Phi_{\lambda m}\cr
F_2(r,\omega,\lambda)\sigma_1\Phi_{\lambda m}\cr}\>, \eqno(3)$$
where $\sigma_1$ is the Pauli matrix, the 2D spinor
$\Phi_{\lambda m}=\Phi_{\lambda m}(\vartheta,\varphi)=
(\Phi_{1\lambda m},\Phi_{2\lambda m})$ is
the eigenspinor
of the twisted euclidean Dirac operator with Chern number $n$ on the unit
sphere with the eigenvalue $\lambda =\pm\sqrt{(l+1)^2-n^2}$ while
$-l\le m\le l+1$, $l\ge|n|$. As for the functions $F_{1,2}$, 
they obey the system of equations
$$\cases{\sqrt{a}\partial_rF_1+
\left(\frac{1}{2}\frac{d\sqrt{a}}{dr}+\frac{\lambda}{r}\right)F_1=
i(\mu_0-c)F_2,\cr
\sqrt{a}\partial_rF_2+
\left(\frac{1}{2}\frac{d\sqrt{a}}{dr}-\frac{\lambda}{r}\right)F_2=
-i(\mu_0+c)F_1\cr} \>\eqno(4)$$
with $c=\omega/\sqrt{a}$ and $a$ of (1). The explicit form of the 2D spinor
$\Phi_{\lambda m}$ is inessential in the given paper and can be found in
Refs.\cite{Gon99}. One can only notice here that
they can be subject to the normalization condition at $n$ fixed
$$\int\limits_0^\pi\,\int\limits_0^{2\pi}(|\Phi_{1\lambda m}|^2+
|\Phi_{2\lambda m}|^2)
\sin\vartheta d\vartheta d\varphi=1$$
and these spinors form an orthonormal basis in $L_2^2({\Bbb S}^2)$ at any
$n\in{\Bbb Z}$.

By passing on to the Regge-Wheeler variable $r_*=r+2M\ln(r/2M-1$) and by
going to the quantities $x=r_*/M, y=r/M, k=\omega M, \mu=\mu_0M$,
we shall have
$x=y+2\ln (0.5y-1)$, so that $y(x)$ is given implicitly by the
latter relation (i.e., $-\infty<x<\infty$, $2\leq y<\infty$) with
$$y'=dy/dx=1-2/y=(y-2)/y=a_0(x)\>\eqno(5)$$
and the system (4) can be rewriten as follows
$$\cases{E'_1 +a_1 E_1=b_1 E_2\>,\cr
 E'_2 + a_2 E_2=b_2E_1\cr} \>\eqno(6)$$
with $E_{1}=E_{1}(x,k,\lambda)=F_+(Mx),
F_+(r^*)=F_{1}[r(r^*)]$,
$E_{2}=E_{2}(x,k,\lambda)=iF_-(Mx),
F_-(r^*)=F_{2}[r(r^*)]$
and
$$a_{1,2}=\frac{1}{2y^2}\pm\frac{\lambda}{y}\sqrt{a_0}\>,\eqno(7) $$
$$b_{1,2}=\mu\sqrt{a_0}\mp k \>.\eqno(8) $$
To evaluate luminosity of the Hawking radiation for spinor particles it is
necessary to know the asymptotics of the functions $E_{1,2}$ at $x\to+\infty$.
As was shown in Ref.\cite{Fir01}, the latter asymptotics look as follows
$$E_1\sim i\sqrt{k-\mu}\,s_{11}(k,\lambda)
e^{ik^+x}e^{i\beta\ln(2k^+x)},\qquad x\to+\infty \>,\eqno(9)$$
$$E_2\sim\sqrt{k+\mu}\,s_{11}(k,\lambda)
e^{ik^+x}e^{i\beta\ln(2k^+x)},\qquad x\to+\infty \>\eqno(10)$$
with $$\beta=\left(\frac{\mu}{k^+}\right)^2\,,\qquad
k^+=\sqrt{k^2-\mu^2}\,,$$
where $s_{11}(k,\lambda)$ is an element of the $S$-matrix connected with
some scattering problem for a Schr\"odinger-like equation (see Section 3).
After this one can obtain (in usual units) the luminosity $L(n)$
with respect to the Hawking radiation for TIC
with the Chern number $n$ in the form
(for more details see Refs.\cite{Gon99,Fir01})
$$L(n)=
A\sum\limits_{\pm\lambda}\sum\limits_{l=|n|}^\infty2(l+1)
\int\limits_{\mu}^\infty\,\frac{|s_{11}(k,\lambda)|^2}
{e^{8\pi k^+}+1}k^+dk=$$
$$A\sum\limits_{l=|n|}^\infty2(l+1)
\int\limits_{\mu}^\infty\,\frac{|s_{11}(k,\lambda)|^2+|s_{11}(k,-\lambda)|^2}
{e^{8\pi k^+}+1}k^+dk\>,\eqno(11)$$
where
$A=\frac{c^5}{\pi GM}\left(\frac{c\hbar}{G}\right)^{1/2}
\approx0.251455\cdot10^{55}\,
{\rm{erg\cdot s^{-1}}}\cdot M^{-1}$
and $M$ in g while $\mu=3.762426569\cdot10^{-18}\mu_0M$
if $\mu_0$ in MeV and $M$ in g.

Luminosity $L(n)$ can be interpreted, as usual,\cite{{GF},{Gon99}} 
as an additional contribution to the Hawking radiation due to the additional
spinor particles leaving the black hole because of the interaction with
monopoles and, as already mentioned in Section 1, the conforming radiation
can be called {\it the monopole Hawking radiation}.\cite{Gon99}

Under this situation,
for the all configurations luminosity $L$ of black hole with respect
to the Hawking radiation concerning the spinor field to be obtained,
one should sum up over all $n$, i. e.
$$L=\sum\limits_{n\in{\Bbb{Z}}}\,L(n)=L(0)+
2\sum\limits_{n=1}^\infty\,L(n)
\eqno(12)$$
since $L(-n)= L(n)$.

It should be emphasized that in (11), generally speaking,
$s_{11}(k,\lambda)\ne s_{11}(k,-\lambda)$. Obviously, for neutrino there 
exists only $L(0)$ since neutrino does not interact with monopoles (when 
neglecting a possible insignificantly small magnetic moment of neutrino).

\section{The scattering problem and description of algorithm}
\subsection{The scattering problem}
   It is evident that for numerical computation of all the above luminosities 
one needs to have some algorithm for calculating $s_{11}(k,\lambda)$. As was
mentioned in Section 2, $s_{11}(k,\lambda)$ is an element of the $S$-matrix 
connected with some scattering problem for a Schr\"odinger-like equation.
The latter equation, as was shown in Ref.\cite{Fir01}, looks as follows 
$$u''+(k^2-\mu^2)u=q_0u \>\eqno(13)$$
with potential
$$q_0(x,k,\lambda,\mu)=\frac{\lambda^2\sqrt{a_0}}{y^2(x)}q(x,k,\lambda,\mu)+
(\sqrt{a_0}-1)\mu^2 \>\eqno(14)$$
while 
$$q(x,k,\lambda,\mu)=\sqrt{a_0}\left(1-\frac{1}{y\lambda^2}\right)+
\frac{y}{\lambda^2}q_1\left(\lambda+\frac{3y}{4\sqrt{a_0}}
q_1\right)-\frac{y^2}{\lambda^2\sqrt{a_0}}\left(\frac{1}{2}q_2+q_3\right)$$
with
$$q_1=-\frac{\mu\sqrt{a_0}}{y^2(k-\mu\sqrt{a_0})}\>,
q_2=-\frac{\mu\sqrt{a_0}(5/y-2)}{y^3(k-\mu\sqrt{a_0})}\>,$$
$$q_3=-\frac{\lambda\sqrt{a_0}}{y^3}\left(y-3+
\frac{\sqrt{a_0}}{\lambda}\right)\>.$$
As an example, in Figs. 1, 2 the numerically computed potentials
$q_0(x,k,\lambda,\mu)$ with $\lambda>0$ and $\lambda<0$ are shown.
It should be emphasized that
potential $q_0$ of (14) satisfies the condition\cite{Fir01} 
$$\int\limits_{-\infty}^{+\infty}|q_0|dx<\infty\>$$
only at $\mu=0$, so when $\mu\ne0$ the correct statement of the scattering
problem for $q_0$ should be modified. Besides one can notice that when 
$\lambda >0$, $q_0(x,\lambda)>0$ at any
$x\in[-\infty,\infty]$ while at $\lambda<0$, $q_0(x,\lambda)$ can change
the sign.

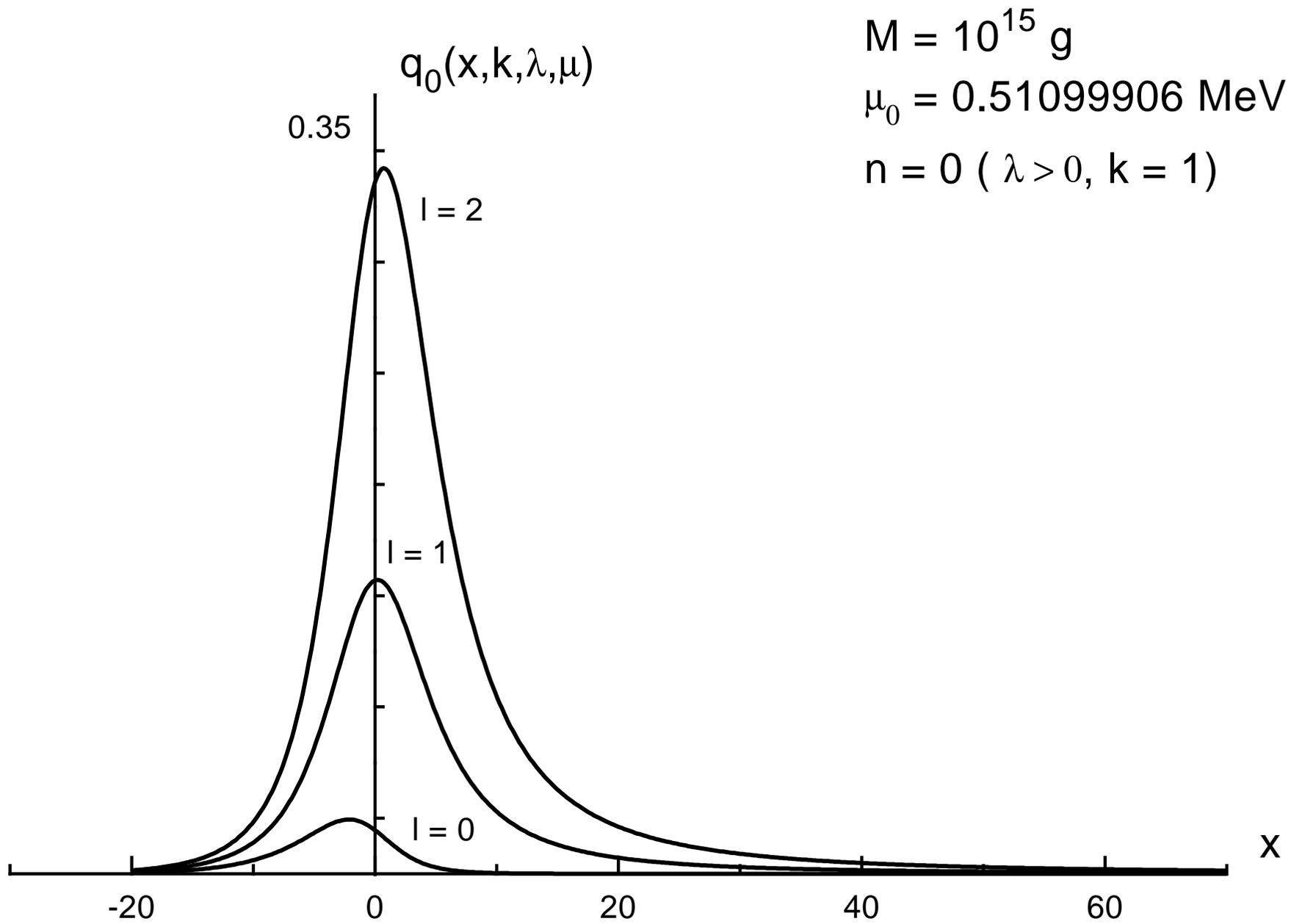
\begin{figure}[th]
\vspace*{8pt}
\caption{Typical potential barrier for $\lambda>0$.}
\end{figure}
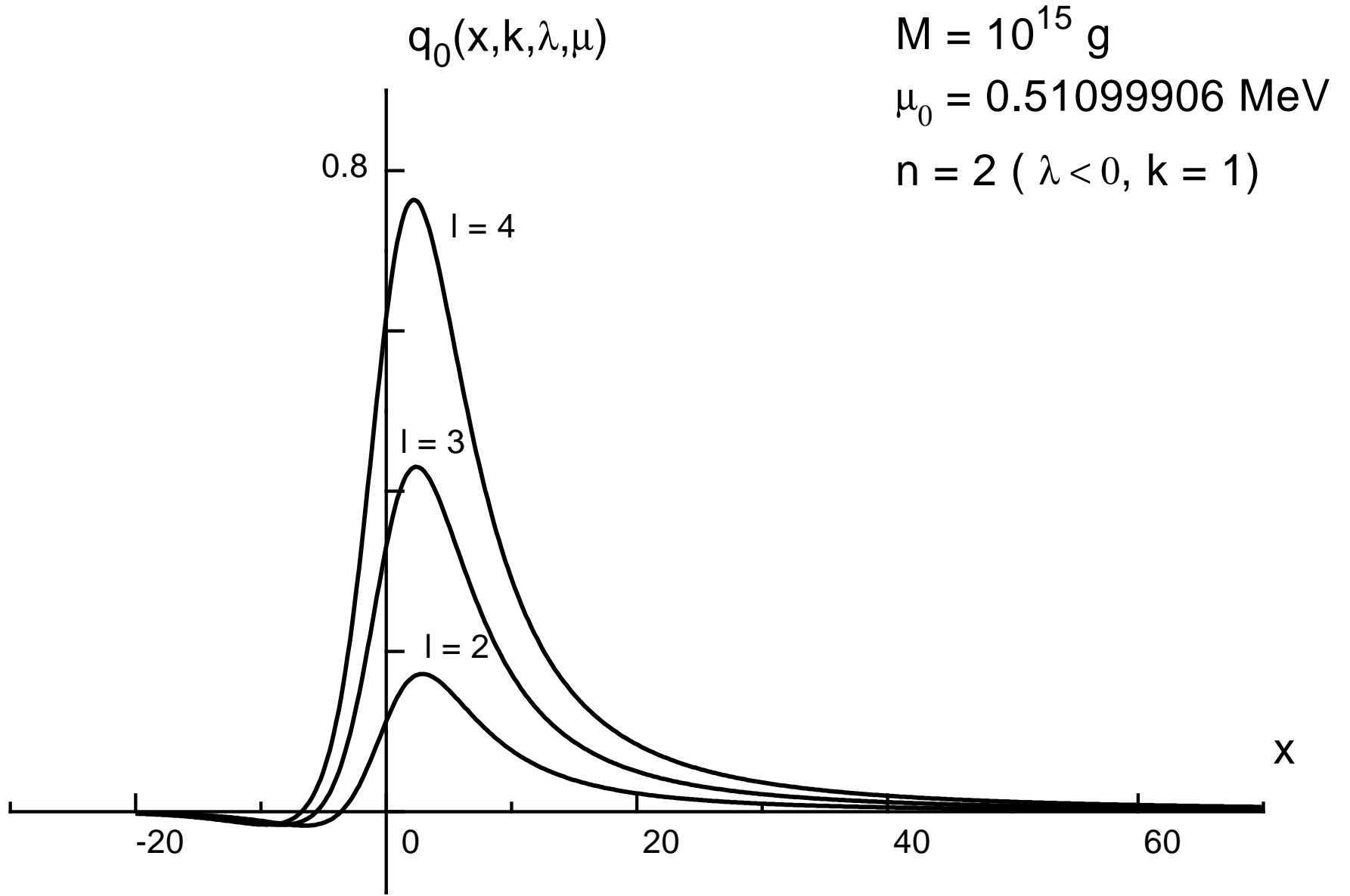
\begin{figure}[th]
\vspace*{8pt}
\caption{Typical potential barrier for $\lambda<0$.}
\end{figure}
In its turn, as was shown in Ref.\cite{Fir01},
the correct statement of the mentioned
scattering problem for Eq. (13) consists in searching for two
solutions $u^+(x,k,\lambda)$,
$u^-(x,k,\lambda)$ of Eq. (13) obeying the following conditions
$$u^+(x,k,\lambda)=
\cases{e^{ikx}+
s_{12}(k,\lambda)e^{-ikx}+o(1),&$x\to-\infty$,\cr
s_{11}(k,\lambda)w_{i\beta,\frac{1}{2}}(-2ik^+x)+
o(1),&$x\to+\infty$,\cr}$$
$$u^-(x,k,\lambda)=\cases{s_{22}(k,\lambda)e^{-ikx}
+o(1),&$x\to-\infty$,\cr
w_{-i\beta,\frac{1}{2}}(2ik^+x)+
s_{21}(k,\lambda)w_{i\beta,\frac{1}{2}}(-2ik^+x)+
o(1),&$x\to+\infty$,\cr}
\eqno(15)$$
where the functions  $w_{\pm i\beta,\frac{1}{2}}(\pm z)$
are related to the Whittaker functions
$W_{\pm i\beta,\frac{1}{2}}(\pm z)$
(concerning the latter ones see e. g. Ref.\cite{AS}) by the relation
$$w_{\pm i\beta,\frac{1}{2}}(\pm z)=
W_{\pm i\beta,\frac{1}{2}}(\pm z)e^{-\pi\beta/2}\>,$$
so that one can easily gain asymptotics (using the corresponding ones for
Whittaker functions\cite{AS})

$$w_{i\beta,\frac{1}{2}}(-2ik^+x)=
e^{ik^+x}e^{i\beta\ln|2k^+x|}[1+O(|k^+x|^{-1})], \> x\to+\infty\>,$$
$$w_{-i\beta,\frac{1}{2}}(2ik^+x)=e^{-ik^+x}
e^{-i\beta\ln|2k^+x|}[1+O(|k^+x|^{-1})], \> x\to+\infty\>.\eqno(16)$$

We can see that there arises some $S$-matrix with elements
$s_{ij}, i, j = 1, 2$.
As has been seen above, for calculating the Hawking
radiation we need the coefficient $s_{11}(k,\lambda)$, consequently, we need 
to have an algorithm for numerical computation of it inasmuch as the latter
cannot be evaluated in exact form. The given algorithm can be extracted from 
the results of Ref.\cite{Fir01}. 
 
\subsection{Description of algorithm}
To be more precise
$$s_{11}(k,\lambda)=2ik/[f^-(x,k,\lambda),f^+(x,k,\lambda)]\>,\eqno(17)$$
where [,] signifies the Wronskian of functions $f^-,f^+$, the
so-called Jost type solutions of Eq. (13).
In their turn, these functions and
their derivatives obey the certain integral equations. Since the Wronskian
does not depend on $x$ one can take the following form of the mentioned
integral equations
$$f^-(x_0,k,\lambda)=e^{-ikx_0}+
\frac{1}{k}\int\limits^{x_0}_{-\infty}
\sin[k(x_0-t)]q^-(t,k,\lambda)f^-(t,k,\lambda)dt\>,\eqno(18)$$
$$(f^-)'_x(x_0,k,\lambda)=
-ike^{-ikx_0}+
\int\limits^{x_0}_{-\infty}\cos[k(x_0-t)]q^-(t,k,\lambda)f^-(t,k,\lambda)dt\>,
\eqno(19)$$
and
$$f^+(x_0,k,\lambda)=w_{i\beta,\frac{1}{2}}(-2ik^+x_0)+$$
$$\frac{1}{k^+}
\int\limits_{x_0}^{+\infty}{\rm Im}[w_{i\beta,\frac{1}{2}}(-2ik^+x_0)
w_{-i\beta,\frac{1}{2}}(2ik^+t)]
q^+(t,k,\lambda)f^+(t,k,\lambda)dt\>,\eqno(20)$$
$$(f^+)'_x(x_0,k,\lambda)=
\frac{d}{dx}w_{i\beta,\frac{1}{2}}(-2ik^+x_0)+$$
$$\frac{1}{k^+}\int\limits^{+\infty}_{x_0}{\rm Im}\left[\frac{d}{dx}
w_{i\beta,\frac{1}{2}}(-2ik^+x_0)
w_{-i\beta,\frac{1}{2}}(2ik^+t)\right]q^+(t,k,\lambda)
f^+(t,k,\lambda)dt\>,\eqno(21)$$
where 
$$q^-(x,k,\lambda)=\frac{\lambda^2\sqrt{a_0}}{y^2}q+
a_0\mu^2\>,
q^+(x,k,\lambda)=\frac{\lambda^2\sqrt{a_0}}{y^2}q+
2\mu^2\frac{y-x}{xy}\>\eqno(22)$$
with $q=q(x,k,\lambda,\mu)$ from (14).

The potential $q^-$ exponentially tends to zero when
$x\to-\infty$ and potential $q^+$ behaves as $O(x^{-2})$ as
$x\to+\infty$. So one can notice that these potentials
are integrable when $x\to-\infty$ or $x\to\infty$ respectively.
The point $x=x_0$ should be chosen from the considerations of
the computational convenience.
The relations (18)--(22) can be employed for numerical calculation of
$s_{11}$. It should be noted, however, that
while Eqs. (18)--(19) are suitable for direct calculation this is not
the case for Eqs. (20)--(21) --- it is difficult to evaluate the functions
$w(z)$ in (20)--(21) since we have yet no effective fast method to compute
the Whittaker functions $W(z)$ at $z$ required. In the present paper we,
therefore, stick to the following strategy. We employ the asymptotic
expressions of $w$-functions at $x\to+\infty$ from Eq. (16) and
replace Eqs. (20)--(21) by

$$f^+(x_0,k,\lambda)=e^{i(k^+x+\mu\ln(2k^+x_0))}+$$
$$\frac{1}{k^+}
\int\limits_{x_0}^{+\infty}{\rm Im}[e^{i(k^+(x_0-t)+\mu\ln(x_0/t))}]
q^+(t,k,\lambda)f^+(t,k,\lambda)dt\>,\eqno(23)$$
$$(f^+)'_x (x_0,k,\lambda)=
ie^{i(k^+x_0+\mu\ln(2k^+x_0))}(k^++\mu/x_0)+$$
$$\int\limits^{+\infty}_{x_0}{\rm Im}\left[i\left(1+\frac{\mu}{k^+x_0}\right)
e^{i(k^+(x_0-t)+\mu\ln(x_0/t))}\right]
q^+(t,k,\lambda)f^+(t,k,\lambda)dt\>.\eqno(24)$$
The latter equations are appropriate for numerical evaluation but under this
approximation, as numerical experiment shows, reliable calculations are only
possible for $\mu< 0.01$. 
This upper limit corresponds to
$\mu_0(e^\pm)$ and $M\sim10^{15}$ g and it is obtained when requiring
that calculations lead to the smooth monotonous graphs for 
$\Gamma(k,\pm\lambda)=|s_{11}(k,\pm\lambda)|^2$
without any oscillations, breaks or fractures (see below Figs. 3,4). In 
its turn, the latter requirement is based on the theoretical results that 
$\Gamma(k,\pm\lambda)$ is a smooth monotonous function of $k$ changing between 
0 and 1 (see Ref.\cite{Fir01}).
Therefore in the present paper we restrict ourselves to considering electrons 
and positrons for $M=10^{15}$ g. Including the heavier fundamental fermions 
($\mu^{\pm}$-mesons and $\tau^{\pm}$-leptons) within the given algorithm is 
only possible for smaller $M$. 

\section{Estimations for luminosities}
Before presentation of the numerical results we should touch upon the 
convergence of the series (11)--(12) over $l$ and $n$ respectively.
For this aim we denote
$$c_l(n)=
\int\limits_{\mu}^\infty\,\frac{|s_{11}(k,\lambda)|^2+
|s_{11}(k,-\lambda)|^2}
{e^{8\pi k^+}+1}k^+dk$$
and represent the coefficients $c_l(n)$ in the form
(omitting the integrand)
$$c_l(n)=c_{l1}(n)+c_{l2}(n)=
\int\limits_\mu^{(\ln l)/2\pi}+\int\limits_{(\ln l)/2\pi}^\infty
\>,\eqno(25)$$
so that $L(n)$ of (11) is equal to $L_1(n)+L_2(n)$ respectively. Also it
should be noted, as follows from the results of Ref.\cite{Fir01}, we
have for barrier transparency $\Gamma(k,\pm\lambda)=|s_{11}(k,\pm\lambda|^2$  
that $\Gamma(\mu,\pm\lambda)=0$ and at $k\to+\infty$
$$\Gamma(k,\pm\lambda)=|s_{11}(k,\pm\lambda)|^2=1+O(k^{-1})\>.\eqno(26)$$
Further we have the asymptotic behaviour for large $l$ and $k<<l$
$$|s_{11}(k,\pm\lambda)|= C\frac{e^{-\sqrt{n+l+1}}}{\sqrt{n+l+1}}
(k-\mu)[1+o(1)]\>
\eqno (27)$$
with some constant $C$. To obtain it we should work using the equations
(18)--(21) by conventional methods
of mathematical analysis so that the detailed derivation of (27) lies somewhat
out of the scope of the present paper and we will not dwell upon it here.
Using (25) and (27) we find
$$c_{l1}(n)\leq B\frac{e^{-2\sqrt{n+l+1}}}{n+l+1}
\int_\mu^{(\ln l)/2\pi}\frac{(k-\mu)^2k^+}{e^{8\pi k^+}+1}dk$$
with some constant $B$ and then it is clear that
$$c_{l1}(n)\leq B\frac{e^{-2\sqrt{n+l+1}}}{n+l+1}
\int_0^{(\ln l)/2\pi}\frac{(k-\mu)^2k^+}{e^{8\pi k^+}-1}dk$$
$$\leq B\frac{e^{-2\sqrt{n+l+1}}}{n+l+1}
\int_0^\infty\frac{(k^+)^3}{e^{8\pi k^+}-1}dk^+=
B\frac{3! \zeta(4)}{(8\pi)^4}\frac{e^{-2\sqrt{n+l+1}}}{n+l+1}\>,$$
where we used the formula (for natural $p>0$)\cite{AS}
$$\int\limits_0^\infty{t^pdt\over e^t-1}=p!\,\zeta(p+1)$$
with the Riemann zeta function $\zeta(s)$, while $\zeta(4)=\pi^4/90$.
As a consequence, one may consider
$$ L_1(n)\sim\int\limits_{n}^{\infty}2(l+1)\frac{e^{-2\sqrt{n+l+1}}}{n+l+1}
dl\sim2\int\limits_{\sqrt{2n+1}}^\infty e^{-2t}\left(t-\frac{n}{t}\right)dt$$
$$\sim e^{-2\sqrt{2n+1}}\left(\sqrt{2n+1}-\frac{1}{\sqrt{2n+1}}\right)
\>, \eqno (28)$$
where we employed the asymptotical behaviour of the incomplete gamma
function\cite{AS}
$$\Gamma(\alpha,x)=\int\limits_x^\infty e^{-t}t^{\alpha-1}dt
\sim x^{\alpha-1}e^{-x},\>x\to+\infty\>,$$
so that the series $\sum_1^\infty L_1(n)$ is evidently convergent.

At the same time due to (26)
$$c_{l2}(n)\leq D\int\limits_{(\ln l)/2\pi}^\infty e^{-8\pi k^+}k^+dk=$$
$$D\int\limits_z^\infty e^{-8\pi k^+}
\frac{(k^+)^2}{k}dk^+\>,\, z=\sqrt{(\ln l/2\pi)^2-\mu^2}\> $$
with some constant $D$.
Since $k^+/k<1$ and 
$$\int\limits_{(\ln l)/2\pi}^\infty e^{-8\pi k^+}k^+dk^+
\sim\frac{\ln l}{l^4}\>$$
then
$$c_{l2}(n)\leq D\frac{\ln l}{l^4}\>. \eqno(29)$$
This entails
$$L_2(n)\sim \int\limits_{n}^\infty\frac{2(l+1)\ln l}{l^4}dl\sim
\frac{\ln n}{n^2}\>.\eqno(30)$$
and the series $\sum_1^\infty L_2(n)$ is also convergent.
Under this situation we obtain that
the series of (12) is convergent, i. e., $L<+\infty$.
Consequently, we can say that all the further computed
luminosities exist and are well defined.

\section{Numerical results}
 In view of (26), we have
$$\int\limits_\mu^{\infty}\frac{\Gamma(k,\lambda)k^+dk}
{e^{8\pi k^+}+1}\sim
\int\limits_0^\infty\,\frac{kdk}{e^{8\pi k}-1}=\frac{1}{(8\pi)^2}\zeta(2)=
\frac{1}{6\cdot8^2}\>,\eqno(31)$$
whilst the latter integral can be accurately evaluated using the
trapezium formula on the [0, 3] interval. Having confined the range of $k$
to the [$\mu$, 3] interval, accordingly, it is actually enough
to restrict oneself to $0\leq l,n\leq$15--20 when computing $L(n),L$.
We took $0\leq l\leq20$, $0\leq n\leq15$.
Since the Wronskian of (17) does not depend on $x$, the latter should be
chosen in the region where the potentials $q^+$ of (22) are already
small enough. Besides,
potentials $q^-$ of (22) are really equal to 0 when $x<$ -20. We computed
$\Gamma(k,\lambda)$ according to (17) at $x=x_0$ = 300,
where $f^\pm$ were obtained
from the Volterra integral equations, respectively, (18) and (23). For this
aim, according to the methods developed for numerical solutions of integral
equations (see e. g., Ref.\cite{Tur87}),
those of (18) and (23) have been replaced by systems of linear algebraic
equations which can be gained when calculating the conforming integrals by
the trapezium formula, respectively, for the [-20, 300] and [300, 400]
intervals. The sought values of $f^\pm$ were obtained as the solutions to the
above linear systems.
After this the derivatives $(f^\pm)'_x$ were evaluated in accordance with
(19) and (24) while employing the values obtained for $f^\pm$. The typical
behaviour of $\Gamma(k,\lambda)$ is presented in Figs. 3, 4.

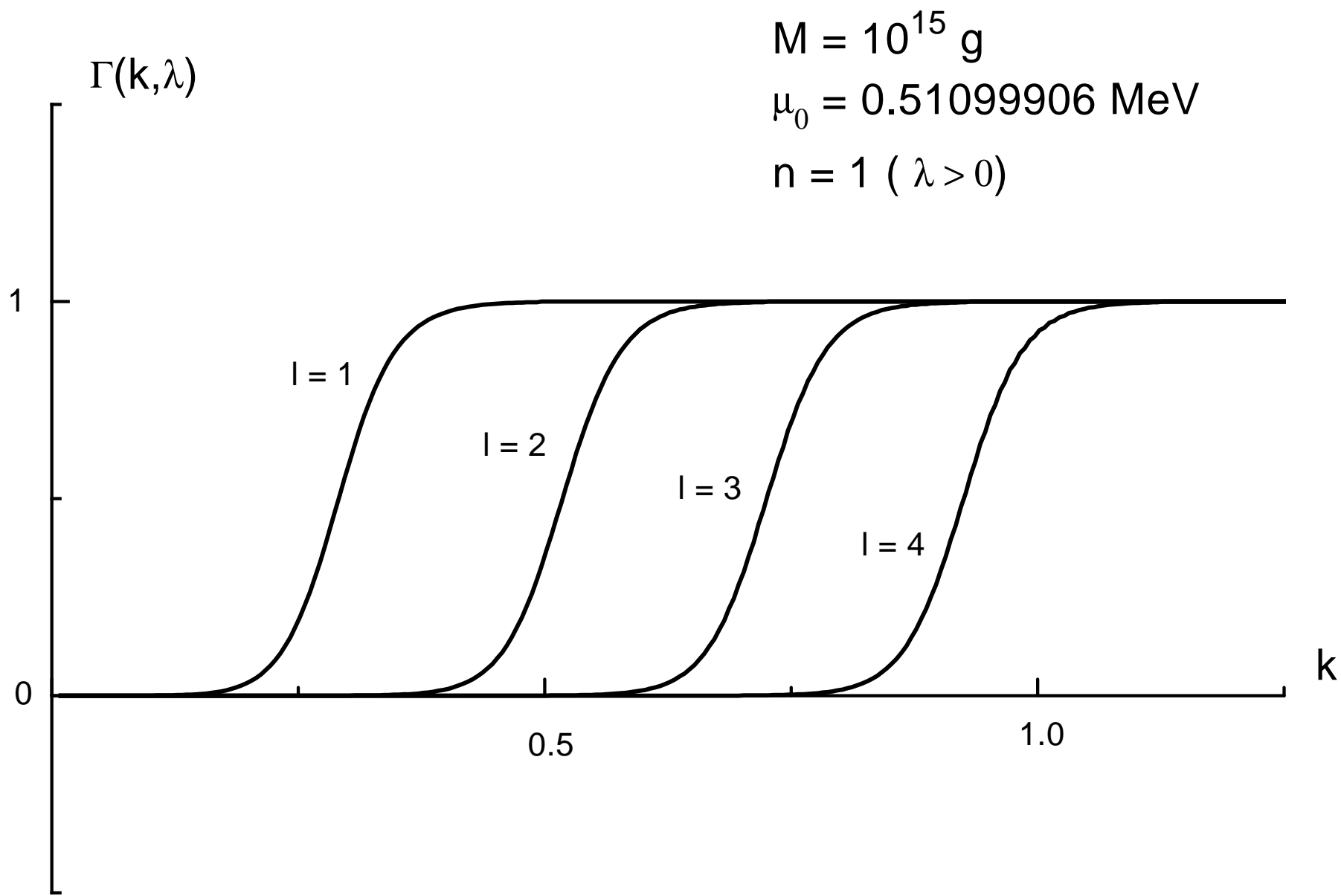
\begin{figure}[th]
\vspace*{8pt}
\caption{Typical behavior of barrier transparencies for $\lambda>0$.}
\end{figure}

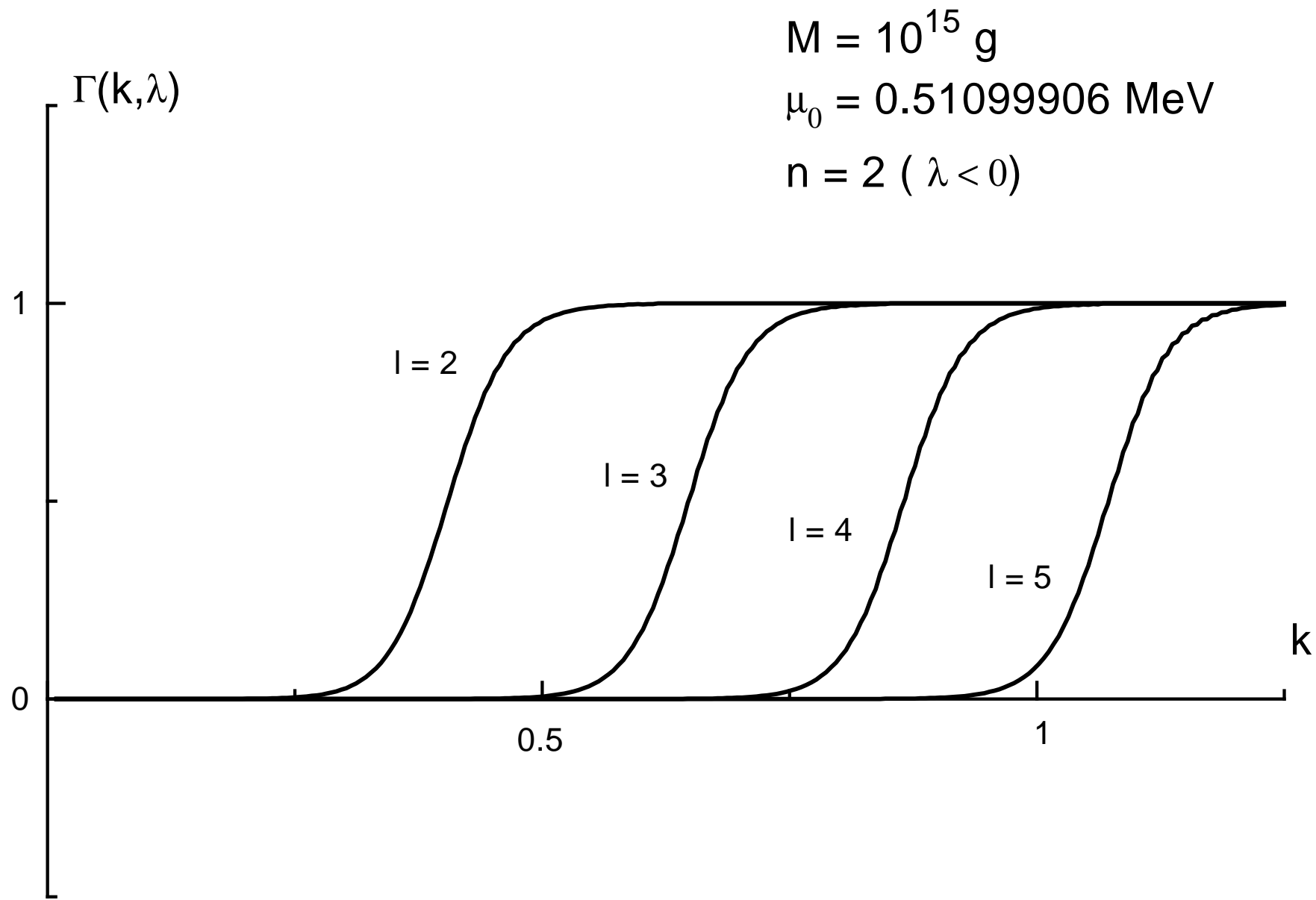
\begin{figure}[th]
\vspace*{8pt}
\caption{Typical behavior of barrier transparencies for $\lambda<0$.}
\end{figure}

Further Fig. 5 presents luminosity $L_{\nu}(0)/A$ with $A$ of (11) for 
electronic neutrino (antineutrino) (no interaction with monopoles) as function 
of $k$ and also the untwisted $L_e(0)/A$ and the all configurations
$L_e/A$ luminosities for electron (positron) as functions
of $k$. The areas under the curves give the corresponding values
of $L(0)/A$ and $L/A$.

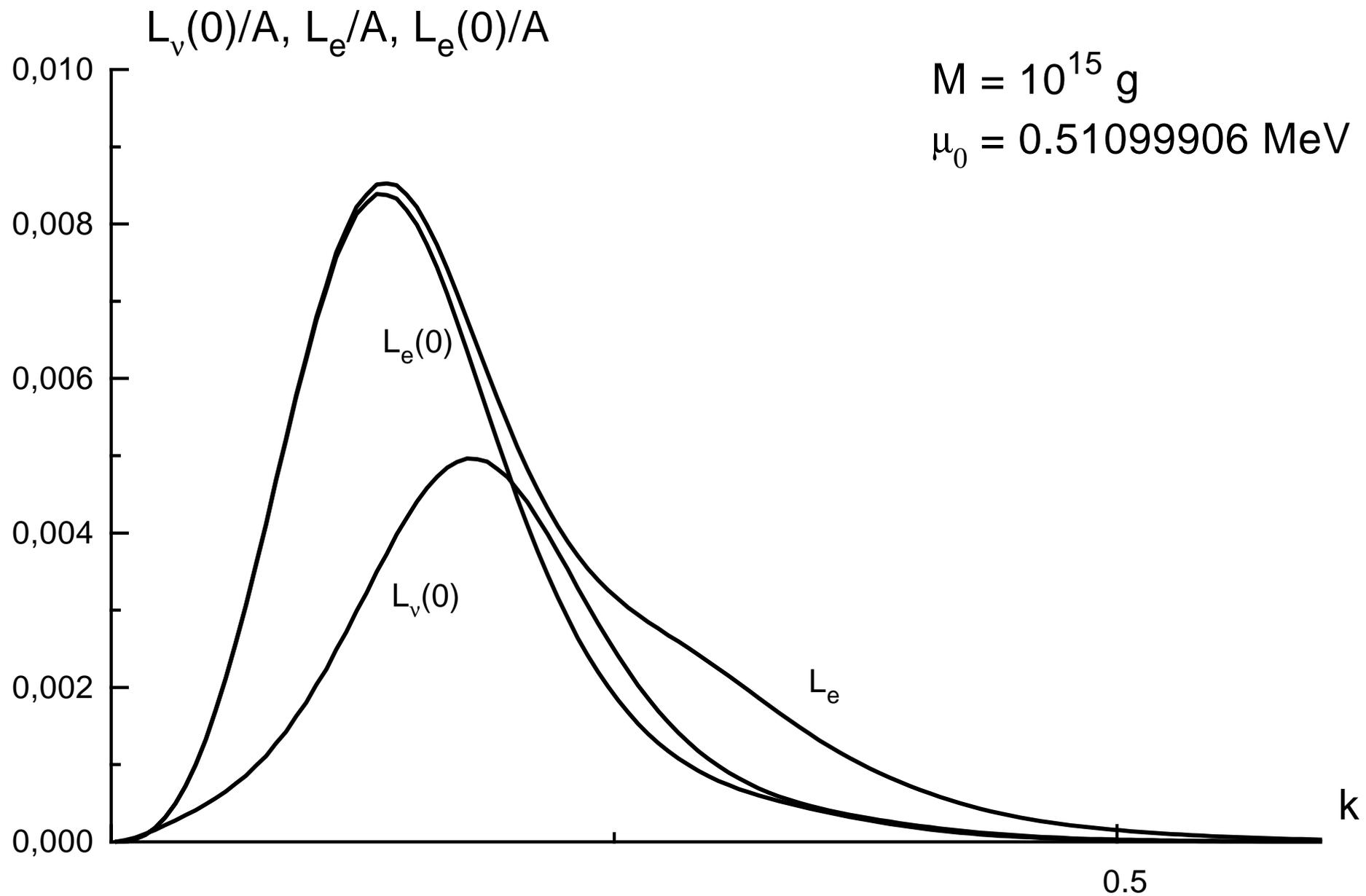
\begin{figure}[th]
\vspace*{8pt}
\caption{Luminosity for electronic neutrino (antineutrino) and untwisted and 
all configurations luminosities for 
electron (positron).}
\end{figure}

\begin{table}[htbp]
\tcaption{Untwisted and the all configurations luminosities.}
{\begin{tabular}{@{}cccc@{}} \\ 
 Particle  & $L(0)/A$ &  $L/A$ &  Monopole contribution of $L$ (\%)  \\
\hline
$\nu_e$ & 0.786745$\cdot10^{-3}$ & 0.786745$\cdot10^{-3}$ &  0\  \\
$\tilde{\nu_e}$ & 0.786745$\cdot10^{-3}$ & 0.786745$\cdot10^{-3}$ &  0\  \\
$e^-$ & 0.120798$\cdot10^{-2}$ & 0.148780$\cdot10^{-2}$  &  18.8072\  \\
$e^+$ & 0.120798$\cdot10^{-2}$ & 0.148780$\cdot10^{-2}$  &  18.8072\  \\  
\hline
\end{tabular}}
\end{table}

  In Table 1 the data on computation of the untwisted and the all
configurations luminosities are represented for all particle species under
consideration.
Finally, we can introduce the {\em total} untwisted ${\cal L}_0$
(as the sum over
all species in the second column of Table 1) and
the {\em total} ${\cal L}$ (as the sum over all species in the third column
of Table 1) luminosities multiplied by $A$
that should be detected by an external observer near black hole. We shall have
  $$ {\cal L}_0 = 0.100317\cdot10^{38}\> {\rm erg\cdot s^{-1}},\>
  {\cal L}= 0.114389\cdot10^{38}\> {\rm erg\cdot s^{-1}}$$
so that contribution owing to Dirac monopoles amounts to 12.3022 \% of
${\cal L}$.

\section{Concluding remarks}
  It seems to us the considerations of the present paper show that exact
quantitative study of the Hawking radiation is not so simple task. At the same
time such an exploration is important enough in a number of astrophysical 
problems, for example, for more precise determination of the masses of
primordial black holes which just expire today. But, as we have seen,
just the black hole mass is one of the crucial parameters that define 
both the relevant scattering problems and the algorithms necessary for 
numerical calculations. Besides the exact evaluations might shed a light on 
nontrivial topological properties of black holes, in particular, on those 
connected with residence of Dirac monopoles (and generally U(N)-monopoles 
with $N>1$) in black holes. As a result, one should continue the exact
calculations within the broader spectrum of both black hole masses and 
quantum particles emitting from black holes, in first turn, for other
fundamental fermions -- $\mu^{\pm}$-mesons and $\tau^{\pm}$-leptons. Also
one should pass on to the other types of black holes, for instance, to the
Reissner-Nordstr\"{o}m ones where the description of the necessary
$S$-matrices for spinor particles has been just recently obtained.\cite{Fir03}

\nonumsection{Acknowledgments}
    The work of Goncharov was supported in part by the Russian
Foundation for Basic Research (the grant No. 01-02-17157).



\vspace*{6pt}
\nonumsection{References}

\end{document}